\documentclass[useAMS,usenatbib]{mnras}
\usepackage{txfonts}
\usepackage{pdflscape}
\usepackage{longtable}  
\usepackage{rotating}
\usepackage{natbib}
\usepackage{graphicx}
\usepackage{graphics}
\usepackage{psfrag}
\usepackage{amssymb}
\usepackage{multicol}
\bibpunct{(}{)}{;}{a}{}{,}
\def\Teff{$T_{\mathrm{eff}}$}
\def\logg{\ensuremath{\log g}}
\def\vmic{$\upsilon_{\mathrm{mic}}$}
\def\vmac{$\upsilon_{\mathrm{macro}}$}
\def\vsini{\ensuremath{{\upsilon}\sin i}}

\def\kms{$\mathrm{km\,s}^{-1}$}

\def\Ro{\ensuremath{R_{\odot}}}

\def\Mo{\ensuremath{M_{\odot}}}

\def\Re{\ensuremath{R_{\oplus}}}
\def\Me{\ensuremath{M_{\oplus}}}

\def\ergscm{erg\,s$^{-1}$\,cm$^{-2}$}
\def\logR{\ensuremath{\log R^{\prime}_{\mathrm{HK}}}}

\def\hd{HD\,219134}

\title[The HD\,219134 multi-planet system I]{Characterisation of the HD 219134 multi-planet system I. Observations of stellar magnetism, wind, and high-energy flux\thanks{Based on observations obtained at the Telescope Bernard Lyot (USR5026) operated by the Observatoire Midi-Pyr\'en\'ees, Universit\'e de Toulouse (Paul Sabatier), Centre National de la Recherche Scientifique of France.}}

\author[C. P. Folsom, et al.]
{C. P. Folsom$^{1}$\thanks{E-mail:colin.folsom@irap.omp.eu},
L. Fossati$^{2}$,
B. E. Wood$^{3}$,
A. G. Sreejith$^{2}$,
P. E. Cubillos$^{2}$,
A. A. Vidotto$^{4}$,
\and
E. Alecian$^{5}$,
V. Girish$^{6}$,
H. Lichtenegger$^{2}$,
J. Murthy$^{7}$,
P. Petit$^{1}$,
G. Valyavin$^{8}$
\\
$^{1}$ IRAP, Universit\'e de Toulouse, CNRS, UPS, CNES, 31400, Toulouse, France\\
$^{2}$ Space Research Institute, Austrian Academy of Sciences, Schmiedlstrasse 6, A-8042 Graz, Austria\\
$^{3}$ Naval Research Laboratory, Space Science Division, Washington, DC 20375, USA\\
$^{4}$ School of Physics, Trinity College Dublin, the University of Dublin, Dublin-2, Ireland\\
$^{5}$ Univ. Grenoble Alpes, IPAG, 38000, Grenoble, France\\
$^{6}$ Space Astronomy Group, ISRO Satellite Centre, Airport Road Bangalore, 560017 India\\
$^{7}$ Indian Institute of Astrophysics, Bangalore 560 034, India\\
$^{8}$ Special Astrophysical Observatory, Laboratory of Stellar Magnetism, Nizhnii Arkhyz, Karachai-Cherkessian Republic, 369167, Russia
}

\begin{document}

\date{}

\pagerange{\pageref{firstpage}--\pageref{lastpage}} \pubyear{}

\maketitle

\label{firstpage}

\begin{abstract}
  HD\,219134 hosts several planets, with seven candidates reported, and the two shortest period planets are rocky (4-5 $M_{\oplus}$) and transit the star. Here we present contemporaneous multi-wavelength observations of the star HD\,219134. We observed HD\,219134 with the Narval spectropolarimeter at the Observatoire du Pic du Midi, and used Zeeman Doppler Imaging to characterise its large-scale stellar magnetic field. We found a weak poloidal magnetic field with an average unsigned strength of 2.5 G. From these data we confidently confirm the rotation period of 42 days, measure a stellar inclination of 77$\pm$8$^{\circ}$, and find evidence for differential rotation. The projected obliquity of the two transiting super-Earths is therefore between 0 and 20$^{\circ}$. We employed HST STIS observations of the Ly$\alpha$ line to derive a stellar wind mass-loss rate of half the solar value ($10^{-14} M_{\odot} {\rm yr}^{-1}$). We further collected photometric transit observations of the closest planet at near-UV wavelengths centred on the Mg{\sc ii}\,h\&k lines with {\it AstroSat}. We found no detectable absorption, setting an upper limit on the transit depth of about 3\%, which rules out the presence of a giant magnesium cloud larger than $9 R_{\rm planet}$. Finally, we estimated the high-energy flux distribution of HD\,219134 as seen by planets b and c. These results present a detailed contemporaneous characterisation of HD\,219134, and provide the ingredients necessary for accurately modelling the high-energy stellar flux, the stellar wind, and their impact on the two shortest-period planets, which will be presented in the second paper of this series.  
\end{abstract}

\begin{keywords}
techniques: polarimetric -- stars: individual: HD\,219134 -- stars: late-type -- stars: magnetic field -- stars: winds, outflows
\end{keywords}
\section{Introduction}\label{sec:intro}
Both radial velocity and photometric transit surveys have revealed the presence of a great diversity in the structure and geometry of planetary systems, and in the physical characteristics of the planets composing them (e.g., \citealt{mullally2015}, \citealt{winn2015}). Although we have already discovered several hundreds of planetary systems, none of them resembles the solar system.

Multi-planetary systems (i.e. systems with more than one planet) allow us to infer more about the formation, structure, and evolution of planets compared to single-planet systems. For example, the presence and detection of gravitational interactions between planets composing a planetary system allows one to measure their masses through radial velocity and/or transit timing variations, and possibly to detect further planets in the system \citep[e.g.,][]{ttv,cloutier2017}. In addition, the presence of mutual inclination between the orbits of planets in the same system gives us clues about their past interaction and future stability of the system \citep[e.g.,][]{veras2004}.

Key parameters allowing us to infer something about the past formation and evolution history of planets and of their orbits are the (sky-projected) obliquity (i.e. the angle between the angular momentum vector of the host star and that of the planetary orbit) and the mutual inclination among the orbits in a planetary system.

On the basis of the solar system, for which the obliquity is of about 6$^{\circ}$, one might expect an average good alignment between the stellar rotation and the orbits of planets. This is in fact generally true, particularly for host stars cooler than 6250\,K, however there are a number of systems that have been found to have significant mis-alignments, even among the cooler host stars \citep[e.g.,][]{queloz2010,winn2010,albrecht2012,bourrier2018}. The exact reasons for these mis-alignments are unknown and a number of theories, connected with either the star, the planets, or neighbouring stars, have been put forward to explain the observations  \citep[e.g.,][]{fabrycky2007,chatterjee2008,batygin2012,rogers2012,spalding2014}.

A low obliquity and an orbital alignment within a given system suggest that the system did not go through strong dynamical interactions and that migration happened quietly within the disc. In contrast, mis-aligned systems and large mutual inclinations among planets in a system are strong indications of the presence of significant planet-planet scattering through the Kozai-Lidov mechanism \citep[e.g.,][]{winn2015}. For closely packed systems of small planets, even small mutual inclinations can potentially lead to orbit crossings, thus planetary collisions \citep{veras2004}.

Measuring obliquities is therefore one of the keys to unravel the history of planetary systems, but it is not straightforward as it requires the knowledge of the planetary orbital and stellar inclination angles. Planetary orbital inclination angles can be derived almost exclusively for transiting planets \citep[see][for the case of a non-transiting planet]{brogi2013} by fitting the transit light curve. Measuring stellar inclination angles is more complicated, though. The most direct method consists of comparing the stellar projected rotational velocity (\vsini), measurable from high-resolution spectra, and the stellar rotational velocity, measurable from the stellar rotation period \citep[e.g., from spot crossings observable in light curves;][]{mcquillan2014} and radius \citep[see e.g.,][]{hirano2012,hirano2014,walkowicz2013,morton2014}. Further extremely powerful methods are Doppler tomography \citep[e.g.,][]{gandolfi2012,zhou2016} and that based on the detection and characterisation of the Rossiter-McLaughlin effect, which describes the stellar rotation along the transit chord, thus giving the possibility to measure the alignment between the orbit of a planet and the stellar rotation axis \citep[e.g.,][]{cegla2016,bourrier2018}. The stellar inclination angle can also be measured by employing the Zeeman Doppler Imaging technique, employed in this work, that can be then combined with the measured orbital inclination of the transiting planets to derive the obliquity.

Planets also allow us to gather precious information about their host stars, for example regarding stellar winds, that would otherwise be impossible to obtain. The winds of late-type stars are optically thin and hence extremely difficult to detect and study. 
Several indirect methods have placed upper limits on these winds, such as the detection of radio \citep[e.g.,][]{fichtinger2017} and X-ray emission \citep[e.g.,][]{wargelin2002}.  Stellar winds can actually be detected and their properties inferred from detections of astrospheric absorption \citep[e.g.,][]{wood2004}.
However, detections of stellar winds through astrospheric absorption are currently available for only a small number of objects. 
The observation and modelling of the interaction between the stellar wind and the extended atmosphere of a planet allows one to instead directly infer the physical properties of the wind (i.e.\ temperature and velocity) at the position of the planet \citep[e.g.,][]{bourrier2013,kislyakova2014,vidotto2017}. The ideal case is that of a multi-planet system in which the stellar wind-planetary atmosphere interaction is detected for more than one planet. This would provide constraints on the stellar wind temperature and velocity at multiple distances from the star, while also knowing that, for example, the stellar wind mass-loss rate is always the same, hence providing a solid anchor on the stellar wind density and velocity. 

Direct magnetic field detections for stars hosting planets are quite rare \citep[e.g.][]{Vidotto2014, Mengel2017}. Magnetic activity makes detecting planets difficult, thus successful planet searches generally are biased towards stars with the least magnetic activity.  Such weak magnetic fields are difficult to detect, requiring long exposures even for very bright stars, thus very few stars have both planet detections and magnetic field measurements.  However, observational knowledge of a star's magnetic field is necessary for understanding how the wind is sculpted by the magnetic field, and thus the impact of the wind on planets.  Given the general lack of stellar wind detections, and paucity of magnetic field measurements for planet hosts, there are very few planet hosting stars for which we have information about both the wind and magnetic field, and until now none for which the observations were contemporaneous. 

The multi-planet system HD\,219134 is key to further advancing our understanding of planets and stellar winds. With five close-in planets (orbital separation $a$ smaller than 0.4\,AU) and one distant gaseous giant planet ($a$\,$\approx$\,3\,AU), HD\,219134 is one of the few systems known to date that very roughly resembles the solar system's architecture, and it lies just 6.5\,pc away from us \citep{motalebi2015,vogt2015}. The six planets orbit a 0.81\,\Mo\ main-sequence K3 star \citep[the stellar radius is 0.778\,\Ro;][]{boyajian2012,gillon2017}, with an estimated age of 11.0$\pm$2.2\,Gyr. The old age is confirmed by the long stellar rotation period of about 40 days and by the low average value of the \logR\ stellar activity parameter of about $-$5.02 \citep{motalebi2015,vogt2015}. For comparison, the basal chromospheric flux level of main-sequence late-type stars is \logR\,=\,$-$5.1 \citep{wright2004} and the average solar \logR\ value is $-$4.902$\pm$0.063 \citep[95\% confidence level;][]{mamajek2008} and ranges between a minimum of about $-$5.0 and a maximum of about $-$4.8 along the solar activity cycle.
HD\,219134 has a 11.7 year chromospheric activity cycle \citep{johnson2016}, similar to the Sun.
A surface average longitudinal magnetic field of 1.1$\pm$0.1\,G was detected for \hd\ \citep{marsden2014}. 

The planets were first detected by radial velocity and, making use of {\it Spitzer} light curves, \citet{motalebi2015} and \citet{gillon2017} discovered that the two innermost planets are transiting. This allowed a precise measurement of the planetary densities, revealing that both planets have an Earth-like density, with less than 10\% uncertainty \citep{gillon2017}.
The inner most planet (b) has a mass, radius, and equilibrium temperature of $4.74 \pm 0.19$ \Me, $1.602 \pm 0.055$ \Re, and 1015 K, while the further out transiting planet (c) has a mass, radius, and equilibrium temperature of $4.36 \pm 0.22$ \Me, $1.511 \pm 0.047$ \Re, and 782 K \citep{gillon2017}.
The outer planets are not transiting, due to their distance from the star, and hence their radii are unknown and the measured masses are lower limits. The rather large density of the two transiting innermost planets suggests the lack of a hydrogen-dominated envelope. This is confirmed by the low value of the restricted Jeans escape parameter $\Lambda$ \citep{fossati2017} of the two planets ($\Lambda$\,$\simeq$\,22 and 28 for \hd\,b and c, respectively), implying that a hydrogen-dominated envelope would have likely escaped within a few hundred Myr \citep{fossati2017}. 
\citet{DornHeng2018} arrived at the same conclusion by employing a Bayesian inference method based on the stellar properties and modelling the escape through the energy-limited approximation. 

\citet{tian2009} showed that super-Earths similar to \hd\,b and c, subjected to hundreds times more high-energy (XUV) stellar flux than the Earth at present, would completely lose their CO$_2$ content within about 1\,Gyr. Considering that the star was more active in the past and that the work of \citet{tian2009} is based on planets in the habitable zone ($\approx$300\,K), the two planets have likely lost most, if not all, of their secondary atmosphere. One can therefore expect that \hd\,b and c have lost both primary, hydrogen-dominated, and secondary, CO$_2$-dominated, atmospheres because of the high temperature and high-energy stellar radiation.

The close proximity to the star of planets a and b, and their lack of a dense atmosphere, implies a dense stellar wind impacting on the planetary surfaces.  It is expected that the surface of both planets sputter atoms and molecules, similar to what occurs on Mercury. The elements released from the planetary surface would build up a thin, metal-rich exosphere \citep{schaefer2009,ito2015}, consisting mostly of Na, O, Si, and Fe atoms/ions \citep[e.g.,][]{miguel2011,kite2016}.  Some of the atoms may dissociate and ionise, and their structure and velocity would then be controlled by the stellar wind properties and the interplanetary magnetic field carried by the stellar wind. The structure of this exosphere is discussed in the second paper in this series (Vidotto et al.\ 2018b).

The high-energy stellar flux and stellar wind therefore play a fundamental role in shaping the evolution of these planets, and in controlling the formation and characteristics of a metal-rich exosphere. In particular, the stellar wind drives the sputtering processes leading to the formation of a metal-rich exosphere, while the XUV flux is mostly responsible for the ionisation processes on-going in the thin exosphere.

The main objective of this paper to place the strongest observational constraints currently possible on the magnetic field, wind, and  XUV flux of \hd.
This is the first in a series of two works aiming at constraining the properties of the stellar wind and XUV flux, and using this to model the planetary exospheres. In this first paper, we present the results of spectropolarimetric, spectroscopic, and photometric observations. We derive the map of the surface magnetic field and analyse the stellar Ly$\alpha$ line to constrain the wind mass-loss rate. We further analyse space-based ultraviolet (UV) observations obtained with {\it AstroSat}. The second paper will then be dedicated to modelling the stellar wind and planetary metal-rich exosphere.

This work is structured as follows. In Sect.~\ref{sec:specparams} we present the spectropolarimetric observations of \hd\ and derive its stellar parameters, and in Sect.~\ref{sec:zdi} we present the results of the Zeeman Doppler Imaging analysis. In Sect.~\ref{sec:astrosphere} we present the analysis of the stellar Ly$\alpha$ emission line of \hd\ and the astrospheric detection. In Sect.~\ref{sec:astrosat} we show the {\it AstroSat} observations, around the transit of \hd\ b, and the resulting upper limit on the detectability of planet b at near-ultraviolet wavelengths.  In Sect.~\ref{sec:xuv}, we derive the high-energy fluxes of the star and high-energy flux estimates at the distance of planets b and c.  In Sect.~\ref{sec:conclusions} we discuss the results of the observations and gather the conclusions.

\section{Stellar spectroscopic parameters}\label{sec:specparams}
We observed HD\,219134 using the high resolution spectropolarimeter Narval \citep{Auriere2003-Narval-early} on the T\'elescope Bernard Lyot (TBL) at the Observatoire du Pic du Midi in France. The instrument contains a cross-dispersed \'echelle spectrograph, fibre-fed from a Cassegrain mounted polarimeter module. It has a spectral resolution of 65\,000 and nearly continuous wavelength coverage from 3700 to 10500 \AA. Observations were obtained in the standard Stokes $V$ mode, which provides Stokes $I$ (total intensity) and Stokes $V$ (circular polarisation) spectra simultaneously.  One observation consists of a sequence of four sub-exposures, with polarimetric optics rotated between each sub-exposure to minimise instrumental effects. The observations were reduced with the {\sc Libre-ESpRIT} pipeline \citep{Donati1997-major}, using a version optimised for Narval, which performs calibrations and optimal spectrum extraction.  The calibration includes bias subtraction, flat fielding, and rejection of bad pixels.  The order geometry is traced using flat field frames, slit orientation is modelled using Fabry-Perot frames, and wavelength calibration is performed using ThAr spectra.  
The optimally extracted spectra for each beam in each sub-exposure are combined to produce a final $I$, $V$, and null test spectra following Eqs.\ 1, 2, and 3 of \citet{Donati1997-major}. 
  
We obtained 15 observations of HD\,219134 from the 20th of June to the 30th of August 2016, spanning 1.7 rotation cycles of the star. These observations are summarised in Table \ref{tab:specpol-obs}.

\begin{table}
\caption[]{Spectropolarimetric observations of HD\,219134 with Narval. Signal to noise ratios are peak values, near 730 nm, for Stokes $V$, per 1.8 \kms\ spectral pixel.  }
\label{tab:specpol-obs}
\begin{center}
\begin{tabular}{cccc}
  \hline
  Date  &  Exp.\ Time (s) & Peak SNR & $B_l$ (G) \\
  \hline
20 Jun 2016 & $840\times4$  & 1404 & $+1.7 \pm 0.3$ \\
28 Jun 2016 & $840\times4$  & 1350 & $-0.4 \pm 0.3$ \\
 3 Jul 2016 & $840\times4$  & 1292 & $-0.1 \pm 0.3$ \\
 6 Jul 2016 & $840\times4$  & 1051 & $+0.8 \pm 0.4$ \\
13 Jul 2016 & $840\times4$  &  399 & $-1.2 \pm 1.0$ \\
15 Jul 2016 & $840\times4$  & 1344 & $-2.1 \pm 0.3$ \\
19 Jul 2016 & $840\times4$  & 1344 & $-3.6 \pm 0.3$ \\
25 Jul 2016 & $840\times4$  & 1373 & $-1.7 \pm 0.3$ \\
28 Jul 2016 & $840\times4$  & 1364 & $-0.1 \pm 0.3$ \\
 3 Aug 2016 & $840\times4$  & 1149 & $+1.2 \pm 0.3$ \\
 8 Aug 2016 & $840\times4$  & 1285 & $-0.6 \pm 0.3$ \\
13 Aug 2016 & $840\times4$  & 1156 & $-1.7 \pm 0.3$ \\
19 Aug 2016 & $840\times4$  & 1081 & $+1.1 \pm 0.4$ \\
24 Aug 2016 & $840\times4$  & 1227 & $+1.5 \pm 0.3$ \\
30 Aug 2016 & $840\times4$  & 1182 & $-2.2 \pm 0.3$ \\
  \hline
\end{tabular}
\end{center}
\end{table}


Spectropolarimetric observations are essential for deriving magnetic geometry, and they also allow us to constrain stellar photospheric parameters.  We used this to determine the effective temperature (\Teff), the surface gravity (\logg), the projected rotational velocity (\vsini), the microturbulence velocity (\vmic), the macroturbulence velocity (\vmac), and metallicity by directly fitting synthetic spectra to an observed spectrum, following \citet{Folsom2016}.
We used the observation from the 28th of July 2016, as it was one of the highest SNR observations.  We found no variability in the Stokes $I$ spectra, as determined by inspecting the spectra and looking for changes in line profile shape or strength beyond the noise, thus the choice of observation has no significant impact on this analysis. 
The spectrum was normalised to the continuum by fitting a low order polynomial (typically five) through carefully selected continuum points. Each spectral order was normalised separately.
Synthetic spectra were generated using the {\sc Zeeman} spectrum synthesis code \citep{Landstreet1988-Zeeman1, Wade2001-zeeman2_etc}, and fit to the observations through a Levenberge Marquard $\chi^2$ minimisation routine \citep{Folsom2016}.  Atomic data were extracted from  the Vienna Atomic Line Database \citep[VALD;][]{Ryabchikova1997-VALD-early, Kupka1999-VALD, Ryabchikova2015-VALD3} and MARCS model atmospheres \citep{Gustafsson2008-MARCS-grid} were used as input.  Particularly discrepant lines were removed from the fit as in \citet{Folsom2016}.  Six spectroscopic windows were fit independently (6000--6100, 6100--6200, 6200--6276, 6310--6400, 6400--6500, 6590--6700\,\AA\ excluding regions with significant telluric lines).  The results of the fits from these windows were averaged to produce the final best fit results, and the standard deviation of these was used as the uncertainty, as in \citet{Folsom2016}.  The results are presented in Table \ref{tab:params-spec}.

The parameters \Teff, \logg, \vmic, and metallicity are well constrained, however there is some degeneracy between \vmac\ and \vsini.  We assumed a radial-tangential form for the macroturbulent broadening \citep[e.g.][]{Gray2005-Photospheres}.  Line broadening dominated by macroturbulence produces a marginally better fit to the observations than line broadening dominated by \vsini.  We consider two cases with \vsini\ $= 0$ and \vmac\ $=0$, to provide upper limits on \vmac\ and \vsini, reported in Table \ref{tab:params-spec}.  Thus the upper limits on \vsini\ and \vmac\ are not independent, and as one value approaches the limit the other value becomes small.  This has little impact on the other parameters, with these two cases varying the other best fit values by much less than the reported uncertainties.  

The \Teff\ we derive is in good agreement with the value of \citet[][\Teff\ $=4699 \pm 16$]{boyajian2012} based on an interferometric radius.  Our \Teff\ also agrees with most of the spectroscopic values from \citet[][\Teff\ = $4820 \pm 61$ -- $4941 \pm 50$ K]{motalebi2015}.  Our \logg\ agrees with the spectroscopic values from \citet[][\logg\ = $4.63 \pm 0.10$]{motalebi2015}, mostly due to our larger formal uncertainties, however our \logg\ is closer to the value inferred from the interferometric radius and mass of \citet[][\logg\ $= 4.54$]{boyajian2012} or \citet[][\logg\ $= 4.55$]{motalebi2015}.
The uncertainties from \citet{motalebi2015} appear to be largely statistical while ours allows for some systematic uncertainties (e.g. uncertainties in atomic data), thus we believe our uncertainty is more realistic for a spectroscopic \logg.
 \citet{motalebi2015} find a \vsini\ ($0.4 \pm 0.5$ \kms) that is consistent with our value.  They adopt a microturbulence of $0.35 \pm 0.19$ \kms, which is smaller than our value, although they comment that the parameter is poorly determined and rely on literature calibrations for the value. 
Our metallicity is consistent with that of \citet[][$\mathrm{[Fe/H]} = 0.11 \pm 0.04$ dex]{motalebi2015}.  Thus our spectroscopic parameters generally agree with the established literature values.  

\begin{table}
\caption[]{Stellar atmospheric parameters derived from direct fitting of the Narval spectra. Limits on \vsini\ and \vmac\ are derived by attributing all line broadening to either \vsini\ or \vmac. }
\label{tab:params-spec}
\begin{center}
\begin{tabular}{lc}
  \hline
\Teff\ (K)     & $4756   \pm 86  $ \\
\logg\         & $4.44   \pm 0.27$ \\
\vsini\ (\kms) & $\leq 1.91      $ \\
\vmic (\kms)   & $0.87   \pm 0.19$ \\
\vmac (\kms)   & $\leq 1.55      $ \\
metallicity    & $0.10   \pm 0.07$ \\
  \hline
\end{tabular}
\end{center}
\end{table}


%
\section{Zeeman Doppler Imaging}\label{sec:zdi}
Understanding the strength and geometry of the large-scale stellar magnetic field is essential for modelling the stellar wind \citep[e.g.][]{Vidotto2015}.  Thus observational constraints on the strength and geometry of the stellar magnetic field are needed to understand the wind of HD\,219134 and its impact on the nearby planets.  

In order to detect magnetic fields through the Zeeman effect in our Narval Stokes $V$ spectra, we used the Least Squares Deconvolution \citep[LSD;][]{Donati1997-major,Kochukhov2010-LSD} technique.  This is a multi-line cross-correlation method that produces a pseudo-average line profile with much a higher SNR.  For LSD, we used a line mask based on atomic data from VALD, using an `extract stellar' request for a star with \Teff\,$= 4500$ K and \logg\,$= 4.5$ and a line depth threshold of 0.1.  The `extract stellar' feature of VALD returns a list of astrophysically relevant lines for a star with a specified model atmosphere, based on approximate calculations of line strength.  Lines blended with tellurics and broad features such as Balmer and Mg{\sc i}\,b lines were identified by manually comparing the line mask to observations and were rejected from the line mask, as were lines to the blue of 500 nm due to lower SNR.  A normalising wavelength of 650 nm and Land\'e factor of 1.195 were used.  This is the same procedure as \citet{Folsom2016} and in fact we use the same mask as their 4500 K line mask.  We detected magnetic fields in 11 out of the 15 observations.  We rejected the observation from the 13th of July from our subsequent analysis due to its low SNR, caused by the presence of thin clouds.  

As an initial characterisation of the magnetic field, we calculated the longitudinal magnetic field ($B_l$, e.g. \citealt{Rees1979-magnetic-cog}) from each observation.  This represents the disk averaged line of sight component of the magnetic field, and can be calculated from the ratio of the first moment of Stokes $V$ to the equivalent width of Stokes $I$.  We followed the same procedure as \citet{Folsom2016}, and the $B_l$ values are reported in Table \ref{tab:specpol-obs}.

To characterise the magnetic field in a more detailed fashion, we used Zeeman Doppler Imaging (ZDI), following the method of \citet{Donati2006-tauSco}.   ZDI is a tomographic method for reconstructing the strength and geometry of the photospheric magnetic field at large scales, using rotationally modulated Stokes $V$ line profiles.  For this we used the ZDI code of \citet{Folsom2018}, which iteratively fits the time series of Stokes $V$ profiles, using the maximum entropy regularised fitting routine of \citet{Skilling1984-max-entropy-regularisation}.  The approach of \citet{Skilling1984-max-entropy-regularisation} is to maximise entropy subject to the constraint that $\chi^2$ remains below a statistically acceptable, user specified, value.  For HD\,219134 we used a Voigt function for the model local line profile (i.e., the line profile emerging from one point on the stellar surface), calculated with the approximation of \citet{Humlicek1982-comp-voigt}.  At this low \vsini\ significant Lorentzian wings are seen in the Stokes $I$ LSD profile, and a Gaussian is a poor approximation.  We used a wavelength of 650 nm and Land\'e factor of 1.195 for our model line.  We modelled the star assuming a homogeneous surface brightness as there was no evidence for line profile variability in Stokes $I$ due to brightness spots.

The \vsini\ of HD\,219134 is poorly constrained, as it is below the spectral resolution of Narval, and is likely smaller than the combination of other line broadening mechanisms.  However, the radius of the star is well constrained from the interferometric observations of \citet[][$0.778 \pm 0.005$ \Ro]{boyajian2012}, and the rotation period is known from \citet[][$42.3 \pm 0.1$ days]{motalebi2015}, which implies an equatorial rotational velocity of $0.931 \pm 0.007$\,\kms.
For the local line profile, the Gaussian and Lorentzian widths and line strength were set by fitting the Stokes $I$ LSD profile, using a $\chi^2$ minimisation procedure.  
Using the inferred equatorial velocity as \vsini, we find a best fit Gaussian width of 1.58\,\kms\  ($1\sigma$ width 1.12\,\kms) and a best fit Lorentzian width of 2.15\,\kms.  If we assume a smaller \vsini\ of 0.5\,\kms\ we find a Gaussian width of 1.66\,\kms\  ($1\sigma$ width 1.17\,\kms) and a Lorentzian width of 2.16\,\kms.  This small change in local line width has a negligible impact on the magnetic map, thus we adopt the former values.

\begin{figure}
\begin{center}
\includegraphics[width=\hsize,clip]{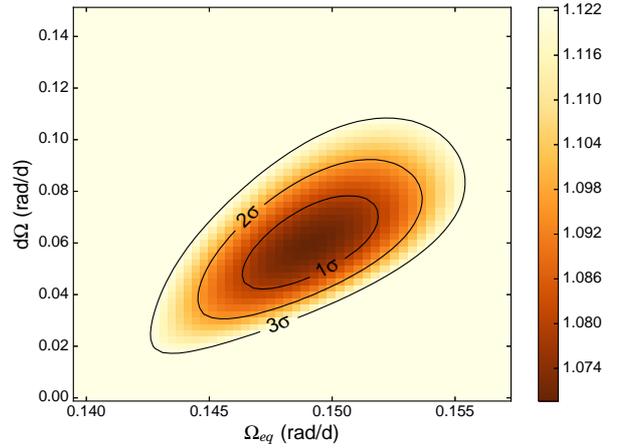}
\caption{Reduced $\chi^2$ as a function of rotation frequency and differential rotation.  Contours indicate confidence levels, based on variations in $\chi^2$ from the minimum.}
\label{fig:period-diffrot} 
\end{center} 
\end{figure}

Some stellar parameters, when they cannot be constrained independently, can be constrained through ZDI modelling, or regular Doppler Imaging (DI).  In particular this can be useful for rotation period, differential rotation, and inclination (and in regular DI \vsini).  The general approach is to search for the model parameters that maximise entropy (minimise information content) for the map \citep[e.g.][]{CameronUnruh1994}.  This can be done simply by running ZDI fits for a grid of model stellar parameters, then selecting the value that maximises entropy.  While this provides an optimal value, it does not provide a formal uncertainty estimate.  In order to derive formal uncertainties knowing the variation in $\chi^2$ around an optimal value is desirable.  The general approach of \citet{Skilling1984-max-entropy-regularisation} can be rephrased to minimising $\chi^2$ subject to entropy remaining above a user specified value \citep[e.g.][]{Petit2002}, provided the user has reasonable target value for entropy.  This image reconstruction at fixed target entropy can then be repeatedly performed for a range of input model parameters, thereby producing the variation in $\chi^2$ around the minimum, for images of constant information content.  In this study we first find the model parameters that maximise entropy for a target $\chi^2$.  Then we use that maximum entropy as a constant target entropy for a grid ZDI models minimising $\chi^2$, and thereby derive the $\chi^2$ landscape around the optimal value.  With this $\chi^2$ landscape, one can then evaluate how statistically significant a change in $\chi^2$ is \citep{Lampton1976-chi2, Avni1976-chi2}, and use that to place formal confidence values on the stellar parameters, as illustrated in Fig.\ \ref{fig:period-diffrot} 

The inclination \hd's rotation axis is not constrained in the literature, and given the large relative uncertainties on \vsini\ measurements they provide no strong constraint.  \hd\,b has an orbital inclination of $85.058 \pm 0.080^\circ$ \citep{motalebi2015,gillon2017} and \hd\,c has an orbital inclination of $87.28 \pm 0.10^\circ$ \citep{gillon2017}, thus if the planets orbits are close to being aligned with the star's rotation axis, the stellar inclination is large.  To constrain the inclination, we used a grid of ZDI models as discussed above, varying the inclination between each by $1^\circ$.  For this search we fixed the equatorial rotational velocity of the model to 0.93\,\kms, but allowed \vsini\ to vary as $i$ changed.  The ZDI model requires an input rotation period and differential rotation (even if it is zero), thus this $i$ estimate depends on these uncertain parameters.  Therefor we adopt an iterative approach to constraining $i$, period, and differential rotation.  We first performed this grid search to get an initial estimate of $i$, then performed the period and differential rotation search discussed below, then repeated this search a second time for a final value of $i = 77^\circ$.  To derive a formal uncertainty, we fit the ZDI model to a constant target entropy in the magnetic map, and allow variations in the resulting $\chi^2$ to place statistical confidence limits on the fits at different $i$.  From this we find $i = 77 \pm 8^\circ$, although this is only a statistical uncertainty and it may be underestimated.  This inclination is within uncertainty of the orbital inclinations of \hd\,b and c.  

Our spectropolarimetric observations span 105 days, and thus can constrain the stellar rotation period.
There is some uncertainty in the rotation period in the literature, with \citet{motalebi2015} finding $42.3 \pm 0.1$ days, and \citet{johnson2016} finding $22.83 \pm 0.03$ but suggesting it may be the first harmonic. 
To investigate this, we ran ZDI models for a grid of rotation periods from 10 to 100 days, and found a clear best fit rotation period of $\sim$45 days.  Thus we conclude that the period of \citet{motalebi2015} is the true value, the period of \citet{johnson2016} is indeed the first harmonic, and the slight discrepancy is due to differential rotation.

\begin{figure}
\begin{center}
\includegraphics[width=\hsize,clip]{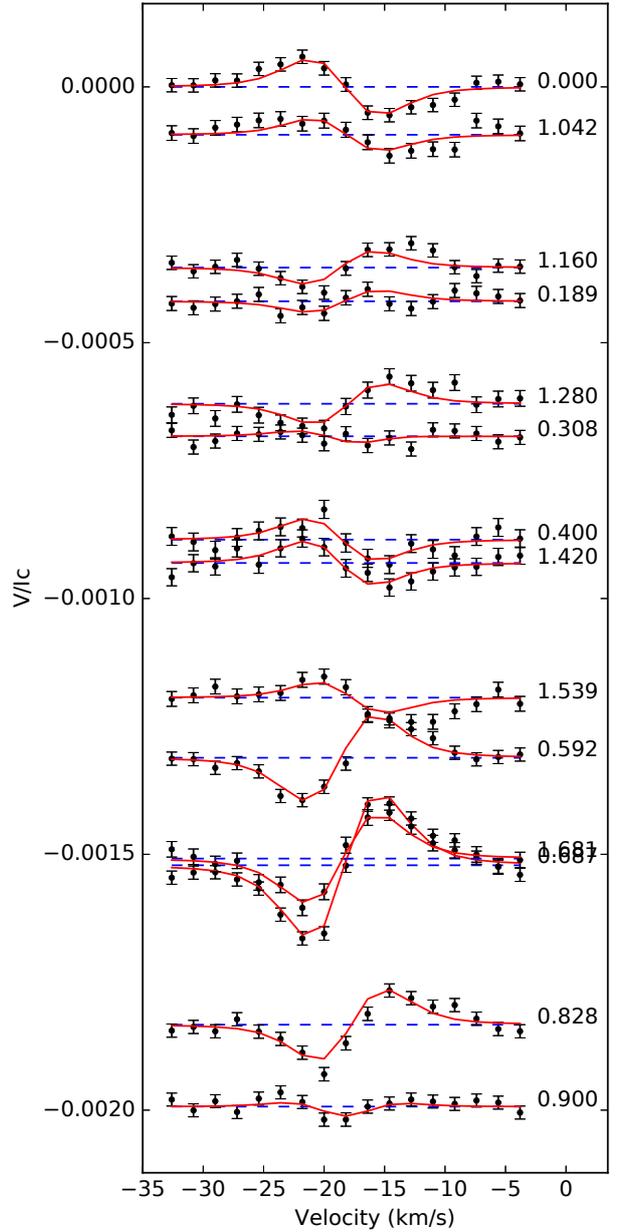}
\caption{Final ZDI fits to the observed Stokes $V$ LSD profiles.  Profiles are shifted vertically according to rotation phase, and labelled by rotation cycle.  Solid lines are the model $V$ profiles and dashed lines represent zero.  }
\label{fig:ZDIfitV} 
\end{center} 
\end{figure}

We found small but significant differences between the observed Stokes $V$ profiles at similar phases separated by one rotation cycle, similar to that at phases 0.3 and 0.5 in Fig.\ref{fig:ZDIfitV}.  With the best fit solid-body rotation rotation period ($\sim$45 days) our models could not reproduce this, (reduced $\chi^2$ 1.2 for 238 degrees of freedom).  This suggests that there is significant evolution of the magnetic field, most likely due to differential rotation, over 105 days.  To account for this in the ZDI model we used the sheared image method developed by \citet{Petit2002}, and a solar-like differential rotation law with the difference in angular frequency between the equator and pole ($d\Omega$) as an additional model parameter (see \citealt{Folsom2018} for this implementation).
We then searched for the optimal combination of rotation period and differential rotation by running ZDI models for a grid of these parameters.  This was iterated with the inclination search discussed above: after, an initial period and $d\Omega$ combination was found we re-derived the best inclination from ZDI, then re-derived the period and $d\Omega$.  Uncertainties were derived by fitting the ZDI models to a constant target entropy and using the different $\chi^2$ values achieved to provide statistical uncertainties.  The resulting $\chi^2$ landscape is plotted in Fig.~\ref{fig:period-diffrot}, with $1\sigma$, $2\sigma$, and $3\sigma$ confidence contours.  The final best fitting values are $P = 42.2 \pm 0.9$\,days and $d\Omega = 0.06 \pm 0.02$\,rad\,day$^{-1}$.  This simple solar-like differential rotation law is sufficient to explain the evolution of the Stokes $V$ profiles over 105\,days, thus we conclude that this is likely the major factor controlling the evolution of the large-scale magnetic field over this time scale.  This differential rotation rate is smaller than the values found by \citet{Folsom2018}, although they focused on younger and somewhat hotter stars than \hd, and their datasets covered a smaller time span.  The differential rotation rate is consistent with that predicted for a 4800\,K star by \citet{Barnes2017}, although their sample is biased towards very rapid rotators.

\begin{figure}
\begin{center}
\includegraphics[width=\hsize,clip]{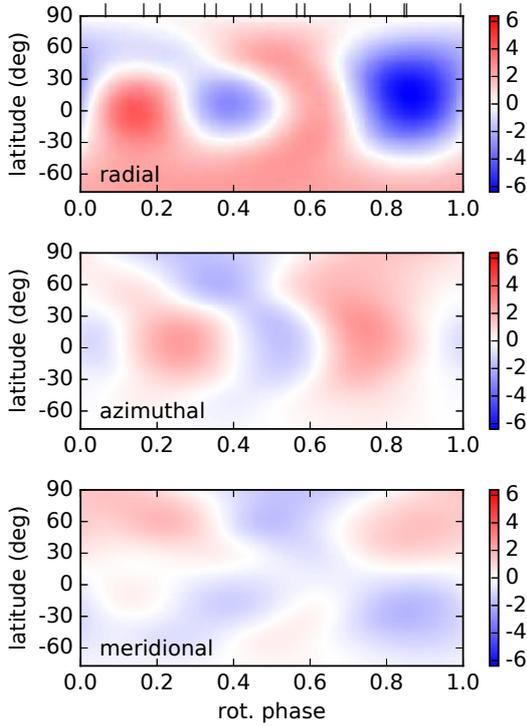}
\caption{2 Final ZDI magnetic map.  The radial, meridional, and azimuthal components of the reconstructed large-scale magnetic field are plotted from top to bottom.  Magnetic field strengths are given in G.  Ticks at the top indicate at which phases the observations were obtained.  }
\label{fig:ZDImap} 
\end{center} 
\end{figure}

The final fits to the observed Stokes $V$ LSD profiles are shown in Fig.~\ref{fig:ZDIfitV} and the magnetic map we derive is presented in Fig.~\ref{fig:ZDImap} (fit to a reduced $\chi^2$ of 0.9). We find the unsigned magnetic field, averaged over the surface of the star, is $\langle B \rangle = 2.5$\,G. We find that 93\% of the total magnetic energy is contained in the poloidal component and only 7\% is toroidal. Of the poloidal magnetic energy, 31\% is in the dipole component, 62\% in the quadrupole, and 4\% in the octupole component.  The total magnetic energy is 20\% symmetric about the rotation axis, and the poloidal component is 19\% axisymmetric (this is summarised in Table \ref{tab:params-mag}).  The large inclination of the rotation axis we find presents a well known limitation of Doppler imaging techniques: when the inclination is large, ZDI (and DI) codes have difficulty determining whether features are in the northern or southern hemisphere of the maps (i.e., there can be an uncertainty in the sign of latitude).
\citet{Auriere2011-EKDra}, in their study of EK Eri, caution that cross-talk may occur between dipole and quadrupole modes when both $i$ is large and \vsini\ is low, as is the case for HD 219134.  However, the magnetic field of HD 219134 is much less axisymmetric than that of EK Eri, making it less vulnerable to such cross-talk.  More over, we clearly see four reversals in the sign of the Stokes $V$ profiles, around phases 0.1, 0.3, 0.5, and 0.9 in Fig.~\ref{fig:ZDIfitV}, which provides clear evidence for a strong quadrupolar component to the field. 
The very low \vsini\ of the star implies that the spatial resolution of our map is low, with much of the resolution provided by phase coverage as features rotate into and out of view.  However, since the lowest degree spherical harmonic components of the poloidal stellar magnetic field control the wind a few stellar radii above the surface \citep[e.g.,][]{Jardine2013, Jardine2017-wind-lowres-mag, See2017-spindown-ZDImaps}, low spatial resolution or an underestimated toroidal field are not serious problems for our study.

\begin{table}
\caption[]{Characterization of large-scale magnetic field from ZDI. Fractions of the magnetic energy in different components are expressed as percentages.  The maximum radial magnetic field strength of the dipolar, quadrupolar, and octupolar components are given. The obliquity of the dipole component with respect to the rotation axis ($\beta_{\rm dip}$) is also given.    }
\label{tab:params-mag}
\begin{center}
\begin{tabular}{lc}
  \hline
  $\langle B \rangle$ (G)  & 2.5 \\
  poloidal (\% tot.) & 93 \\
  toroidal (\% tot.)  & 7 \\
  dipole (\% pol.)  &  31 \\
  quadrupole (\% pol.)  & 62 \\
  octupole (\% pol.)  &  4 \\
  axisymmetric (\% tot.) & 20 \\
  axisymmetric (\% pol.) & 19 \\
  $B_{\rm dip, max}$ (G)  & 2.4 \\
  $B_{\rm quad, max}$ (G) & 4.0 \\
  $B_{\rm oct, max}$ (G)  & 1.2 \\
  $\beta_{\rm dip}$ ($^\circ$)  & 52  \\
  \hline
\end{tabular}
\end{center}
\end{table}


%
\section{Astrospheric Wind Measurement}\label{sec:astrosphere}
\begin{figure*}
\begin{center}
\includegraphics[width=\hsize,clip]{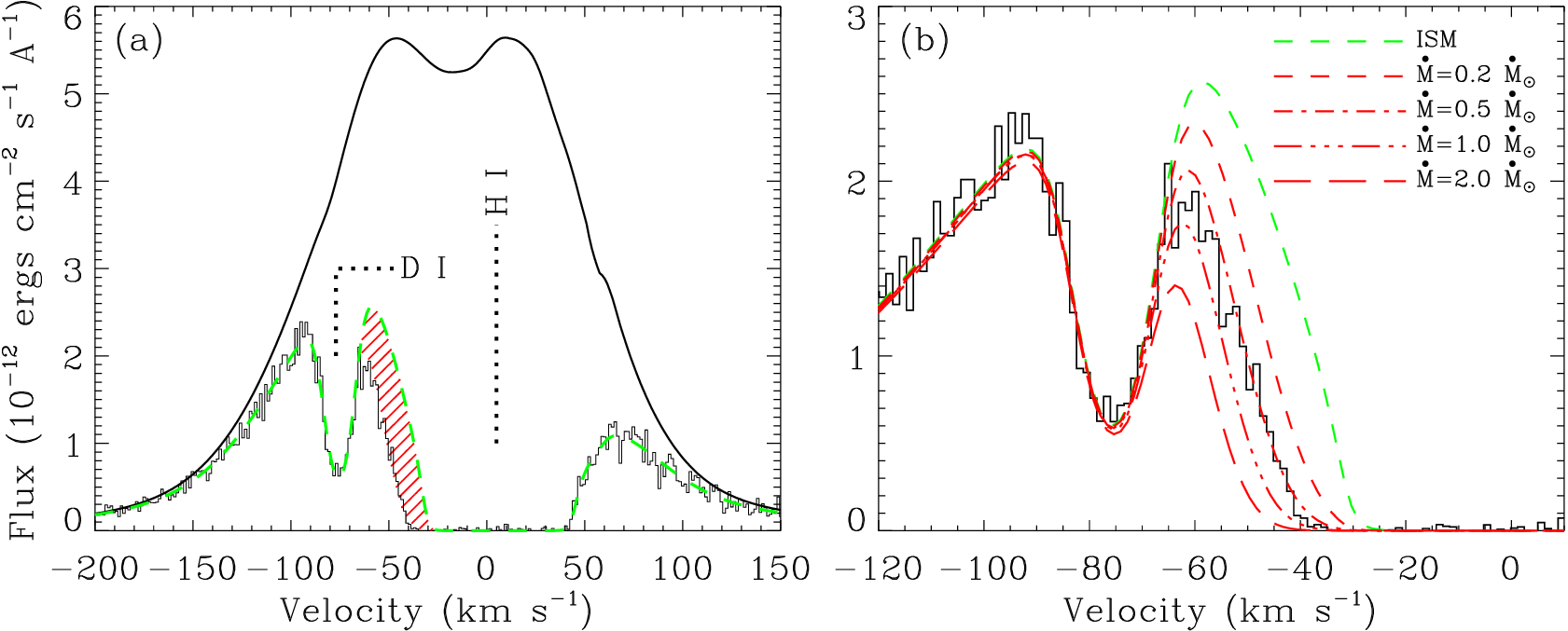}
\caption{(a) The HST/STIS spectrum of the H{\sc i} Ly$\alpha$ line of HD\,219134, plotted on a heliocentric velocity scale, showing absorption from ISM H{\sc i} and D{\sc i} superposed on the chromospheric emission line. The upper solid line is the reconstructed intrinsic stellar emission line profile.  The green dashed line is a fit to the data indicating the ISM absorption observed towards HD\,219134, which fits the data everywhere but on the left side of the H{\sc i} absorption. The excess absorption there (red shaded region) is circumstellar astrospheric absorption. (b) A close-up of the Ly$\alpha$ spectrum near the astrospheric absorption. The predicted absorption of four models of the astrosphere are shown, assuming different stellar mass-loss rates of $\dot{M}=0.2-2.0$\,$\dot{M}_{\odot}$, with $\dot{M}=0.5$\,$\dot{M}_{\odot}$ providing the best fit to the data (where $\dot{M}_{\odot}$ is the solar mass-loss rate $2\times10^{-14} M_{\odot}\,{\rm yr}^{-1}$).}
\label{fig:astrosphere} 
\end{center} 
\end{figure*}

On the 15th of October 2016, HST observed the $1163-1357$\,\AA\ spectrum of HD\,219134 for 1792\,s using the E140H grating of the STIS spectrometer. This spectrum includes the H{\sc i} Ly$\alpha$ line at 1216\,\AA, which is shown in Fig.~\ref{fig:astrosphere}. These data can be used to search for Ly$\alpha$ absorption from the stellar wind/ISM interaction region (i.e., the star's ``astrosphere''), which can be used to estimate the strength of the stellar wind \citep{wood2005a,wood2005b}.

The chromospheric Ly$\alpha$ emission line is greatly affected by absorption from ISM H{\sc i} and D{\sc i} (deuterium) absorption, as is the case for all stars, even those as nearby as HD\,219134. Following the methodology described in detail in \citet{wood2005b}, we first try to fit the observed Ly$\alpha$ absorption assuming that it is entirely from the ISM. In these fits, we force the H{\sc i} and D{\sc i} absorption to be self-consistent, meaning the central velocities of the absorption are forced to be identical (e.g., $v({\rm DI})=v({\rm HI})$), the Doppler broadening parameters are related by $b({\rm DI})=b({\rm HI})/\sqrt{2}$ (appropriate since thermal broadening dominates the Ly$\alpha$ lines in the very local ISM), and we assume an abundance ratio of ${\rm D/H}=1.56 \times 10^{-5}$, which is known to be appropriate for the very local ISM \citep{wood2004apj}. Best fits are determined by minimising the $\chi^2$ statistic \citep{bevington1992}.

We find that we cannot fit the data with only ISM absorption. The H{\sc i} absorption is blueshifted relative to D{\sc i}, and the H{\sc i} absorption is broader than the D{\sc i} absorption would predict. In short, there is excess absorption on the left side of the broad H{\sc i} absorption line that cannot be accounted for by the ISM. This is generally interpreted as a signature of astrospheric absorption \citep{wood2005b}. Fig.~\ref{fig:astrosphere}(a) shows a fit to the data with the left side of the H{\sc i} absorption ignored, providing an indication of the amount of astrospheric absorption present. The best fit ISM parameters of this fit are $v({\rm HI})=7.3\pm0.2$ \,\kms,
$b({\rm HI})=11.2\pm0.2$\,\kms, and column density (in cm$^{-2}$ units) $\log N({\rm HI})=18.03\pm0.01$. The $v({\rm HI})$ and $b({\rm HI})$ values agree reasonably well with the ``Local Interstellar Cloud'' (LIC) predictions of $v({\rm LIC})=5.5\pm1.4$\,\kms\ and $b({\rm HI})=11.24\pm0.96$\,\kms\ from \citet{redfield2008}. The latter prediction is based on the LIC temperature and non-thermal velocity of $T=7500\pm1300$\,K and $\xi=1.62\pm0.75$\,\kms\ 
respectively, with $b({\rm HI})^2=0.0165T+\xi^2$.

It is the astrospheric absorption signature that is of primary interest here, as it is useful as a diagnostic of the stellar wind. The wind strength can be estimated from the absorption with the assistance of hydrodynamic models of the stellar astrosphere. This requires knowledge of the ISM flow vector in the rest frame of HD\,219134. Assuming the LIC vector from \citet{redfield2008}, and the known proper motion and radial velocity of the star (taken from SIMBAD), we find that HD\,219134 sees an ISM wind speed of $v({\rm ISM})=48.5$\,\kms\ with a line of sight from the star towards the Sun that is $\theta=59.9^{\circ}$ from the upwind direction. Rather than compute new astrospheric models specifically for HD\,219134, we note that the $v({\rm ISM})=48.5$\,\kms\ value is close to the $v({\rm ISM})=45$\,\kms\ speed used in previous models of the astrosphere of EV\,Lac, so we instead simply use the existing EV\,Lac models from \citet{wood2005a} to approximate HD\,219134.

The predicted absorption from four of these models is computed for the appropriate $\theta=59.9^{\circ}$ line of sight, and in Fig.~\ref{fig:astrosphere}(b) is compared with the observed astrospheric absorption towards HD\,219134. These models assume various stellar mass-loss rates in the range of $\dot{M}=0.2-2.0$\,$\dot{M}_{\odot}$, where $\dot{M}_{\odot}=2\times 10^{-14}$\,$M_{\odot}$\,yr$^{-1}$ is the solar mass-loss rate. The model with $\dot{M}=0.5$\,$\dot{M}_{\odot}$ clearly provides the best fit to the data, so $\dot{M}=0.5$\,$\dot{M}_{\odot}$ is our best estimate for the wind of HD\,219134. With a stellar radius of 0.778\,\Ro\ \citep{boyajian2012} and an X-ray luminosity (in ergs\,s$^{-1}$) of $\log L_X=26.85$ \citep{schmitt2004}, HD\,219134 has a coronal X-ray surface flux only about a factor of two lower than the Sun's $F_X=3.2\times 10^4$\,\ergscm. Thus, HD\,219134 has a similar level of coronal activity as the Sun. With a stellar surface area roughly half that of the Sun, a wind that is roughly half as strong is therefore what we might have expected, consistent with the astrospheric measurement.
\section{{\it \bf AstroSat} ultraviolet observations}\label{sec:astrosat}
We observed \hd\ with {\it AstroSat} \citep{astrosat} simultaneously employing the twin of 38\,cm Ultraviolet Imaging Telescopes \citep[UVIT;][]{uvit}.
The observations were performed on the 22nd of December 2016 and spanned about 3.7\,hours, divided into three portions each covering about 15, 12, and 8\,minutes. The gaps are due to Earth occultations and instrument overheads. The observations were conducted during the very early phases of {\it AstroSat} science operations, while there were still scheduling problems and hence unfortunately the data were not well centred around the primary transit of \hd\,b (Fig.~\ref{fig:uvit_nuv}). As a result, only the first observation was obtained during the primary transit of \hd\,b, while the second and third observations occurred right after the transit. The data do not cover the transit of \hd\,c.

Since {\it AstroSat} is a relatively new facility, and this is the first dataset of exoplanet transit observations conducted with AstroSat, these data are a good test-bed to assess the possibility of using this space observatory for exoplanet science. This is particularly important because {\it AstroSat} was not designed to detect the small signals typical of exoplanet transit observations. This is highly desirable in light of the limited resources currently available, and foreseen in the near future, to observe at UV wavelengths. Furthermore, the satellite would be, in principle, capable of simultaneously obtaining optical, UV, and X-ray data of a target, which has been shown to be extremely useful for interpreting exoplanet transit observations, thanks to the possibility of disentangling stellar and planetary signals \citep[e.g.,][]{lecav2012}.
\subsection{Ultraviolet and optical photometry}\label{sec:uv}
The UVIT data were obtained in photometric mode\footnote{UVIT observations can be conducted employing either filters or grisms, the latter for low resolution spectroscopy.} employing the neutral density filter in the optical channel, the NUVN2 filter in the near-UV (NUV) channel, which is centred on the Mg{\sc ii}\,h\&k resonance lines (275--283\,nm), and the CaF2 filter in the far-UV (FUV) channel, which is the FUV filter with the broadest available transmission function (for more details, see the UVIT instrument webpage: {\tt http://uvit.iiap.res.in/}). All detectors were employed in photon-counting mode (the equivalent of TIME-TAG mode available on the COS and STIS instruments on-board of HST), allowing us to chose the integration time over which to do the photometry. The star is not detected in the FUV channel and, despite the neutral density filter, the data are strongly affected by leakage in the optical channel, thus making these data unusable. We, therefore, concentrate on the NUV channel.
\begin{figure}
\begin{center}
\includegraphics[width=\hsize,clip]{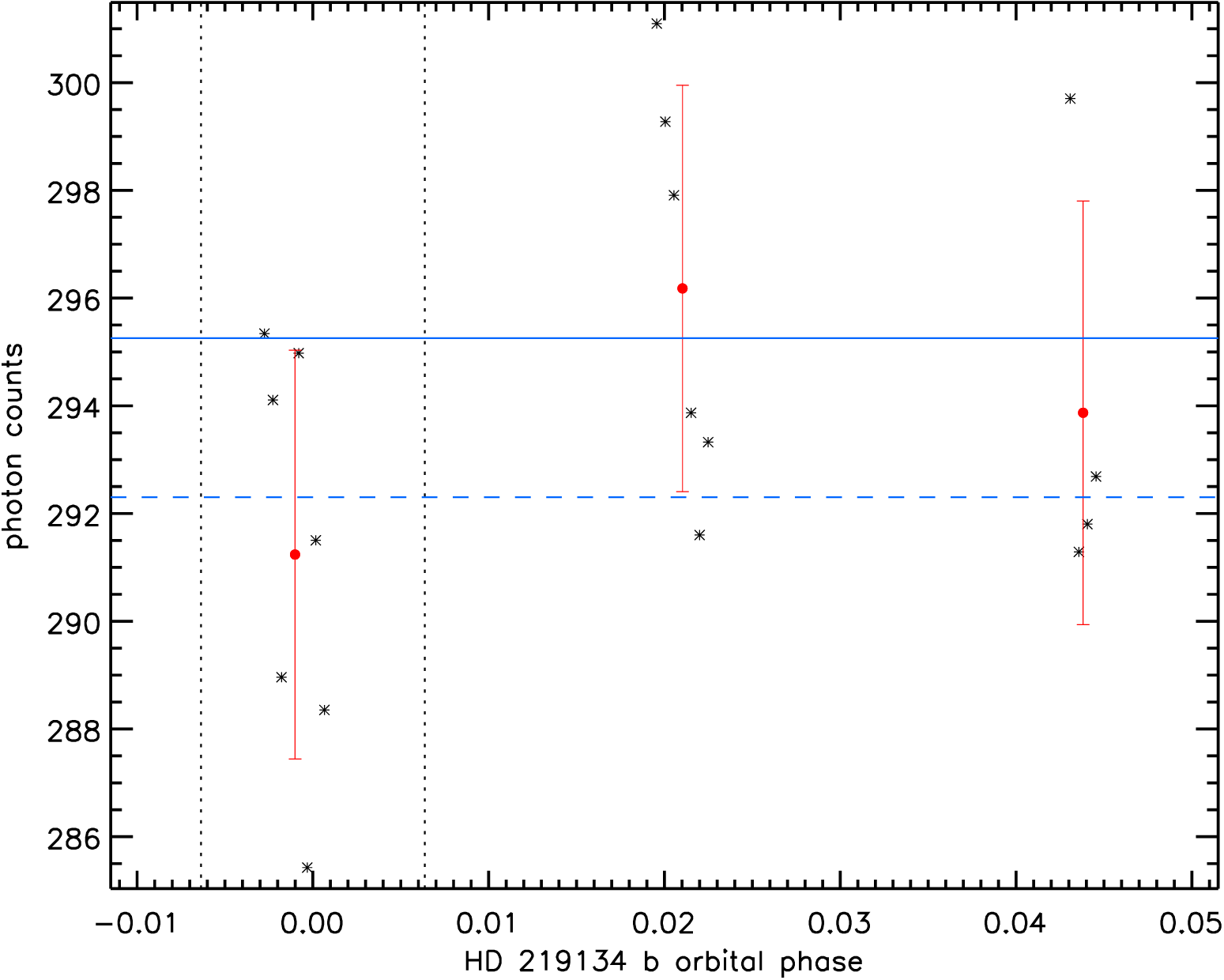}
\caption{UVIT NUV light curve obtained with two minute exposure times (black asterisks). The red dots and their uncertainties mark the average photon count and standard deviation for each observation. The vertical dotted lines indicate the phases of first and fourth contacts according to the optical transit \citep{gillon2017}. The blue solid line marks the average counts for the out of transit observations, while the blue dashed line is set 1\% below the solid line to guide the eye (the optical transit depth is 0.0358\%).}
\label{fig:uvit_nuv} 
\end{center} 
\end{figure}

The satellite is affected by a rather unstable pointing, which makes the target wobble around the centre of the detector with an amplitude of several tens of pixels. The data reduction pipeline we employed, thoroughly described in \citet{murthy2017} and \citet{rahna2017}, corrects for this pointing jitter. We further employed the BLISS-mapping tool as described in \citet{cubillos2013} to check for the presence of residual pixel-to-pixel sensitivity variations without finding any.

Thanks to the photon counting mode, we extracted the number of counts within a circle having a diameter of six pixels around the star using different exposure times, ranging from one second to five minutes. Figure~\ref{fig:uvit_nuv} shows the UVIT NUV light curve obtained with an exposure time of two minutes. The light curve reveals that the counts obtained during the first exposure (i.e., in transit for planet b) are slightly below those obtained out of transit, but the difference is not significant.  From the size of the error bars shown in Figure~\ref{fig:uvit_nuv}, one can infer an upper limit to the NUV transit depth of about 3--4\%.  The depth of the optical transit is 0.0358\%. Since these observations were conducted employing a narrow-band filter centred on the Mg{\sc ii}\,h\&k resonance lines, we can exclude the presence of a large cloud of ionised magnesium surrounding the planet, with an upper limit on the size of 9\,$R_{\rm planet}$ ($R_{\rm planet} = 1.60$ \Re).  In Paper II, our modelling suggests that planet b may host a cloud of ionised magnesium, although nearly all of such a cloud would be within less than $\sim$3\,$R_{\rm planet}$.

\section{High-energy stellar flux}\label{sec:xuv}
On the basis of our HST/STIS observations and of the X-ray flux \citep{schmitt2004}, we derived the XUV stellar flux in three different ways. We first employed the empirical relations given by \citet{linsky2014} and the integrated flux of the reconstructed Ly$\alpha$ line of 9.74$\times$10$^{-13}$\,\ergscm. The results are listed in the third column of Table~\ref{tab:XUVfluxes}. In the 100--912\,\AA\ wavelength range, we obtained an integrated stellar flux at the distance of the \hd\,b and c planets of 496 and 175\,\ergscm, respectively.

\begin{table}
\caption[]{X-ray and EUV fluxes at 1\,AU (in \ergscm) derived for \hd\ by rescaling the solar spectrum (column two) and by employing the scaling relations of \citet[][column three]{linsky2014}, and for comparison from the Sun (column four). The fluxes are calculated in the different bands adopted by \citet{ribas2005} and \citet{linsky2014}.}
\label{tab:XUVfluxes}
\begin{center}
\begin{tabular}{l|ccc}
\hline
Wavelength  & \multicolumn{3}{|c}{Flux at 1\,AU} \\
Range [\AA] & \multicolumn{3}{|c}{[\ergscm]}     \\
            & \multicolumn{2}{|c}{\hd}    & Sun 	      \\
\hline
1--20                   & 0.002 &	 & 0.013 \\
20--100                 & 0.030 &	 & 0.150 \\
100--360                & 0.243 &	 & 1.228 \\
360--920                & 0.157 &	 & 0.794 \\
920--1180               & 0.110 &	 & 0.553 \\
1--360$+$920--1180      & 0.526 &	 & 2.657 \\
1--1180                 & 0.543 &	 & 2.738 \\
\hline
100--200                & 0.089 &  0.095 & 0.451 \\
200--300                & 0.055 &  0.106 & 0.276 \\
300--400                & 0.109 &  0.233 & 0.548 \\
400--500                & 0.016 &  0.010 & 0.079 \\
500--600                & 0.026 &  0.020 & 0.133 \\
600--700                & 0.022 &  0.025 & 0.112 \\
700--800                & 0.023 &  0.031 & 0.115 \\
800--912                & 0.056 &  0.043 & 0.283 \\
912--1170               & 0.100 &  0.183 & 0.503 \\
\hline
\end{tabular}
\end{center}
\end{table}


Because of the similarities in coronal activity between the Sun and \hd\  (Sect.~\ref{sec:astrosphere}), we rescaled the solar irradiance reference spectrum \citep{woods2009} to match the reconstructed Ly$\alpha$ flux, accounting for \hd's distance and radius. We then integrated the rescaled solar flux within the X-ray and EUV wavelength ranges adopted by \citet[][their Table~4]{ribas2005} and \citet[][their Table~5]{linsky2014}. The results are shown in the second column of Table~\ref{tab:XUVfluxes}. The EUV flux obtained in this way is about a factor of two smaller than that inferred from the scaling relations of \citet{linsky2014}.  The X-ray flux (0.517--12.4\,nm; see Sect.~\ref{sec:astrosphere}) is a factor of five larger than the observed one.  These are within the accepted uncertainties \citep{linsky2014}.

Finally, we employed the scaling relation provided by \citet{chadney2015} to convert the observed X-ray flux into an EUV flux. We obtained an EUV flux at one AU of 1.705\,\ergscm, which implies values of 1135 and 400\,\ergscm\ for \hd\,b and c, respectively. These values are about a factor of two larger than those obtained from the scaling relations of \citet{linsky2014}, which is also within the expected uncertainties.
\section{Conclusions}\label{sec:conclusions}
\hd\ is a bright star hosting several planets and due to the very close, transiting orbits of \hd\,b and c, is a prime candidate for studying the impact of stellar radiation and winds on rocky planets.  In this study we have characterised the strength and geometry of the large-scale photospheric magnetic field of the star, derived the wind mass-loss rate of the star, and calculated the EUV and X-ray flux of the star.  Our spectropolarimetric, UV spectroscopic and photometric observations were obtained within six months, thus the derived quantities are contemporaneous.  This detailed characterisation of the star provides the necessary ingredients for a uniquely detailed modelling of the stellar wind, and thus of the impact of the stellar radiation and wind on the planets, which will be presented in a subsequent paper.

Very few stars with known exoplanets also have known magnetic field strengths and geometries, here we add \hd\ to that number. 
We find that \hd\ has an average unsigned large-scale magnetic field strength of 2.5\,G, making it among the weakest main sequence stars with a reliably detected magnetic field \citep[e.g.][]{Vidotto2014}.  This is comparable to the large-scale (low degree spherical harmonic) magnetic field seen in the sun \citep[e.g.,][]{Vidotto2016}.  This weak field is consistent with the very long rotation period of the star, and \hd\ is one of the most slowly rotating main sequence stars for which a magnetic map has been derived.  The dominantly poloidal magnetic field we derived is similar to that found in other very slowly rotating stars \citep[e.g.,][]{Petit2008, Folsom2016}.  However the quadrupole component being stronger than the dipole component of the field (62\% and 31\% of the poloidal magnetic energy, respectively) is rare among both fast and slowly rotating G and K stars (e.g., \mbox{\citealt{Petit2008}}, \mbox{\citealt{Rosen2016}} \mbox{\citealt{Folsom2018}}). 
The solar quadrupolar field becomes stronger than the dipolar magnetic field when the sun is near activity maximum (\citealt{DeRosa2012}; \citealt{Vidotto2018-sun-mag}).  According to the chromospheric activity cycle of \citet{johnson2016} \hd\ was over half way from minimum to its activity maximum during our observations, but likely did not reach maximum for another 2 years.  The weak differential rotation we find of $d\Omega = 0.06 \pm 0.02$\,rad\,day$^{-1}$ is in the range expected for a star of this temperature, and this value being low is essential for our ability to map a star with such a long rotation period. 

As a by-product of the ZDI analysis, we measured a stellar inclination angle of 77$\pm$8$^{\circ}$. The two transiting super-Earths, \hd\,b and c have measured orbital inclination angles of 85.05$\pm$0.09$^{\circ}$ and 87.28$\pm$0.10$^{\circ}$, respectively \citep{gillon2017}, which leads to obliquities of 8$\pm$8$^{\circ}$ and 10$\pm$8$^{\circ}$, respectively. The projected obliquity is therefore small (in the solar system the obliquity is about 6$^{\circ}$), which suggests a smooth disc migration with gentle planet-planet scattering. This is not too surprising considering that this is an old system, and one of the richest known in terms of number of planets. Therefore, the long-term stability of such a large number of planets is likely due to weak planet-planet interactions.

Detections of stellar winds and measurements of mass-loss rates for cool main sequence stars remain very rare.  To our knowledge, \hd\ represents the first cool main sequence star with a contemporaneous mass-loss rate measurement and magnetic field map.  The value we derive on the basis of an astrospheric detection ($\dot{M}=0.5$\,$\dot{M}_{\odot}$ or $10^{-14}\,M_{\odot}\,{\rm yr}^{-1}$) is consistent with what suggested by the stellar X-ray flux and radius, and comparable to the solar mass-loss rate allowing for the difference in radius.  The contemporaneous mass-loss rate and large-scale photospheric magnetic field will allow us to make the most accurate model possible of the 3D stellar wind. 

We further employed {\sc AstroSat} NUV photometry to attempt the detection of the primary transit of the most close-in planet, but without success. This sets an upper limit of about 9\,$R_{\rm planet}$ on any extended cloud of gas as seen at the wavelength of the Mg{\sc ii}\,h\&k resonance lines. Our analysis also suggests that {\it AstroSat/UVIT} is capable of providing NUV photometry allowing only the detection of signals significantly larger than 1\%, at least for something as bright at HD\,219134.

Finally, we employed three different and independent methods to estimate the stellar XUV flux, which drives atmospheric escape. The three methods agree within a factor of about five for EUV flux values integrated over the 100--912\,\AA\ wavelength range, with the best values producing about 500 and 200\,\ergscm\ at the distance of the \hd\,b and c planets, respectively.

In the accompanying paper, we present a 3D MHD model of the stellar wind, and then model the impact of the stellar wind and high-energy flux on the two inner-most rocky planets.  We further model the stellar wind induced sputtering to describe the formation and geometry of the possible metal-rich exosphere surrounding \hd\,b and c. 

\section*{Acknowledgements} 
We are grateful to Joshi Santosh for having provided the link towards obtaining {\it AstroSat} observations.
AAV and CPF acknowledge joint funding received from the Irish Research Council and Campus France through the Ulysses funding scheme.  GV thanks the Russian Science Foundation (project No. 14-50-00043, "Exoplanets") for support of his participation in the analysis of the Astrosat UV-observations.

\bsp

\label{lastpage}

\end{document}